\documentstyle[a4,12pt,epsf]{article}

\begin{document}

\newcounter{bild}
\font\menge msbm10 at12pt
\newcommand{\biltit}[2]{\nopagebreak \centerline{Figure \ref{#1}{\sl #2}}}
\newcommand{\av}{\mathop{\rm av}}
\newcommand{\A}{\cal A}
\newcommand{\D}{\cal D}

\title{Perfect 3-Dimensional Lattice Actions\\
for 4-Dimensional Quantum Field\\
Theories at Finite Temperature }
\author{U. Kerres, G. Mack \\ 
II. Institut f\"ur Theoretische Physik der Universit\"at Hamburg, \\
Luruper Chaussee 149, 22761 Hamburg, Germany
\and G. Palma \\ 
Departamento de Fisica, Universidad de Santiago de Chile,\\
Casilla 307, Correo 2 , Santiago, Chile}
\maketitle

\begin{abstract}
We propose a two-step procedure to study the order of phase transitions
at finite temperature in electroweak theory and in simplified models
thereof.

In a first step a coarse grained free energy is computed by perturbative
methods. It is obtained in the form of a 3-dimensional perfect lattice
action by a block spin transformation. It has finite temperature
dependent coefficients. In this way the UV-problem and the infrared
problem is separated in a clean way. In the second step the effective
3-dimensional lattice theory is treated in a nonperturbative way, either
by the Feynman-Bogoliubov method (solution of a gap equation),
by real space renormalization group methods,
or by computer simulations.

In this paper we outline the principles for $\varphi ^4$-theory and
scalar electrodynamics. The Ba{\l}aban-Jaffe block spin transformation for
the gauge field is used. It is known how to extend this transformation
to the nonabelian case, but this will
not be discussed here.
\end{abstract}

\newpage

\section{Introduction}
\label{secintro}
The possibility of restoration of  spontaneously broken symmetry in the
electroweak theory at high temperature  has recently led to a renewed
interest in the electroweak phase transition \cite{kapusta}.

Since the seminal and pioneering paper by Kirzhnits and Linde
\cite{kirzhnits},
considerable effort has been devoted to describe the rather involved
processes occurring very close to the critical temperature.
The effective potential has shown great usefulness.
It gives the free energy as a function of the magnetization.
There are perturbative computations
combined with $1/N$ expansions for the effective potential \cite{dolan}, and
also
numerical results based on Monte Carlo simulations and on three dimensional
reduced actions \cite{baig}.

Nevertheless the perturbative calculations have shown to be plagued with
problems which are to a large extent due to infrared divergences and are
manifested in the appearance of spurious complex terms in the expansion for
the effective potential. A great amount of work has been done to obtain
meaningful expressions very close to the critical surface. Most of these
attempts are related to the resummation of an infinite number of daisy and
superdaisy diagrams which have to be taken into account in a consistent
perturbative expansion, because they are of the same order of magnitude in
the coupling constants \cite{takahashi}.

Another very successful approach was proposed by
Buchm\"uller et al. \cite{buchmuller},
who use an improved perturbation expansion for the effective potential,
where the dynamically generated plasma masses are already included in the
corresponding expressions from the beginning. These plasma masses are
computed up to one loop order by self-consistent gap-equations, and have the
effect of damping the infrared divergences. In this context, in
\cite{hebecker} the
absence of  linear terms in the scalar field of the effective potential was
explicitly verified for the abelian Higgs model, up to the order $e^4$ and $%
\lambda ^2$, and proved to remain valid at higher orders. Such a linear term
contains spurious infrared divergent contributions.

Despite the important and promising contributions to the understanding of
the nature of the phase transition that have already been made, there are
still ambiguities to be explained and higher order corrections to be
included consistently to ensure the survival of the predictions to all
orders.

We propose an alternative variational two-step method to study phase
transitions at finite temperature in the electroweak theory, and in related
models. It consists first in the calculation of a coarse grained free energy
by perturbative methods, obtaining a 3-dimensional perfect lattice action as
a function of a block spin field.
Their  coefficients depend explicitly on the temperature . In a second
step, the Feynman-Bogoliubov method can be used with this 3-dimensional lattice
action to obtain the best quadratic approximation
to the perfect action, and therefore the best split into a free part
and an interaction.
Using this, the constraint effective potential can be computed.
To lowest order, the masses are obtained in this way as solutions of a
gap equation whose formal solution is the sum of superdaisy diagrams
for the 3-dimensional lattice theory.
They depend on temperature.
Higher order corrections can be computed in principle.
These masses
are to be inserted into the expression for the constraint effective
potential.

This procedure separates the UV- and IR-problems in a clean
way.
It allows to study the perfect 3-dimensional lattice action directly
by numerical simulations.
One could for instance compute the constraint effective
potential numerically.

The paper is organized as follows. In section 3 the block spin
transformation for the scalar fields at zero temperature is discussed. The
evaluation of the perfect scalar action at T=0 is explained and the
extension to finite temperature is outlined.
An explicit
computation of the leading terms of the perfect action for the $
\lambda \Phi ^4-$theory is shown here as well.
In section 4 the Feynman-Bogoliubov method is
applied to the lattice action of the previous section and a comparison with
other approaches is performed here and in appendix \ref{dimred}.
In section 5 we extend the procedure to the Maxwell theory at zero temperature.
In particular, the averaging operator for abelian gauge fields, the perfect
lattice action, the gauge fixing and the interpolation kernel are discussed.
The block spin of Ba{\l}aban and Jaffe for the abelian gauge field  is
used \cite{balaban}.
In section 6 the procedure is extended to scalar electrodynamics.
Here the main features of the extension of the perfect lattice
action to finite temperature are outlined and some quantities which
are needed for perturbative calculations are exhibited.
A short version of the results of this paper was presented at LATTICE 94
in Bielefeld \cite{bielefeld}.

\section{Scholium : Free propagators at finite temperature}
\label{secscholium}

To prepare for the block spin transformation at finite temperature, we
recall first the well known relation between free propagators at zero
temperature and at finite temperature $T=\beta ^{-1}$

Because of translation invariance, the standard free propagators in
Euclidian space time depend
on a single variable $x=({\bf r},t)$. The zero temperature propagator
has the form
\begin{eqnarray}
v_0({\bf r},t)=\int \frac{d^3{\bf p}}{(2\pi )^3}\frac{d\omega }{2\pi }
\exp (-i {\bf p}\cdot {\bf r}-i\omega t){ \left[ \omega ^2+{\bf p}^2
+m^2 \right] } ^{-1}
\end{eqnarray}

A finite temperature field theory lives on a Euclidean space time which
is periodic in time direction with period $\beta $, i.e. on a tube.
Propagators admit a well known random walk
representation. The difference between zero and finite temperature stems
from the possibility that these random walks may wind around the
tube several times. Accordingly, the finite temperature propagator
$v_T$ is obtained from $v_0$ by periodizing in time.

\begin{eqnarray}
v_T({\bf r},t)=\sum_{n\in {\menge Z}}v_0({\bf r},t+n\beta )
\end{eqnarray}
One may use Poisson's resummation formula.

\begin{eqnarray}
\sum_{n \in {\menge Z}}
\exp (-i\omega (t+n\beta ))=\exp (-i\omega t)\sum_{n\in
{\menge Z}}2\pi \beta ^{-1}\delta (\omega -\omega
_n),\,\,\,\, \mbox{with} \,\,\,\,\omega _n=2\pi n\beta ^{-1}
\end{eqnarray}
As a result , the well known finite temperature propagator is obtained
as a sum over Matsubara frequencies $\omega _n$
\begin{eqnarray}
v_T({\bf r},t)=\sum_{n \in {\menge Z}}\beta ^{-1}\int \frac{d^3{\bf p}}{(2\pi
)^3} \exp (-i{\bf p}\cdot {\bf r}-i\omega_n t)\left[ \omega _n^2+{\bf
p} ^2+m^2\right] ^{-1}
\end{eqnarray}
Later on, we will concentrate on the $m=0$ case.

\section{Scalar fields only}
\label{secscalar}

To introduce our method we start with a theory with
scalar fields only.
We consider first the $T=0$ case and define the  blockspin,
the effective action, the interpolation operator and the
fluctuation field propagator.
Then we deal with the $T > 0$ case i.e. with the generalizations of
the above mentioned entities.
As an illustration we write down the effective action for $T > 0$
to leading order in the fluctuation field propagator.
Finally we add a note on the cancellation of the $T$-dependent UV
divergencies.

\subsection{Block spins for scalar fields at zero
temperature}
\label{subsecblock}

We start from a scalar field theory on the continuum ${\menge R}^4$of points z.
The continuum is divided into blocks $x$. Normally one chooses
hypercubes, but here we wish to admit different extensions $L_s$
of the blocks in space direction and $L_t$ in time
direction. We identify the blocks $x$ with the sites at their centers.
In this way a lattice $\Lambda $ of lattice spacing ($L_s,L_t$)
in space and time direction is obtained. With a scalar field
$\varphi (z)$ one associates a block spin $ \Phi (x)$ . Following
Gawedzki and Kupiainien \cite{gawedzki}, we choose them as block averages.

\begin{eqnarray}
\Phi (x)=C\varphi (x)=\av_{z \in x}\varphi (z)
\end{eqnarray}
$C$ is called the averaging operator. It has a kernel $C(x,z)$ which
equals the properly normalized characteristic function $\chi
_x(z)$ of the block $x.$ Because of invariance under simultaneous
translations of $z$ and $ x$ by lattice vectors in $\Lambda $, its
Fourier representation has the form
\begin{eqnarray}
C(x,z) &=&
\frac 1 {L_s^3 L_t} \chi_x(z)
\nonumber \\
&=&  \sum_l\int \frac{d^4p}{(2\pi )^4}\exp
(-ip(z-x)-ilz)\widetilde{C}(l,p)
\label{ctilde}
\end{eqnarray}
Summation over $l=({\bf l},l_4)$ runs over $l_4\ \in $ 2$\pi
L_t^{-1}{\menge Z}$ and ${\bf l}$ $\in$ (2$\pi L_s^{-1}{\menge Z}$)$^3.$
Integration over p is over the first Brillouin zone.
\begin{eqnarray}
\parallel {\bf p}\parallel \leq \pi L_s^{-1} \,\,\,\,\, , \,\,\,\,\,\,
\mid p_4\mid \leq \pi L_t^{-1}
\end{eqnarray}
Explicitly
\begin{eqnarray}
\widetilde{C}(l,p) &=& \widetilde{C}_t(l_4,p_4)
\widetilde{C}_s ({\bf l}, {\bf p})
\nonumber \\
\widetilde{C}_t(l_4,p_4) &=& \frac 2{L_t(p_4+l_4)}\sin
\frac{ L_t(p_4+l_4)}2
\nonumber \\
\widetilde{C}_s ({\bf l}, {\bf p})
 &=& \prod_{j=1}^3\frac 2{L_s(p_j+l_j)}\sin
\frac{L_s(p_j+l_j)}2
\end{eqnarray}

\subsection{Definition of the perfect action}

Given the continuum action $L$($\varphi $) , the perfect 4-dimensional
lattice action $L_\Lambda $($\Phi $) at zero temperature is defined
in Wilson's way
\begin{eqnarray}
\exp (- L_\Lambda (\Phi )) = \int D\varphi \mbox{ }\delta
 (C\varphi -\Phi) \exp (- L(\varphi))
\end{eqnarray}
In place of the $\delta $-function, one may use a Gaussian
\begin{eqnarray}
\delta _\kappa (C\varphi -\Phi) = \prod_{x \in \Lambda}
(\frac{2\pi }\kappa)^{-\frac 12} \exp (-\frac
12\kappa(C\varphi (x)-\Phi (x))^2)
\end{eqnarray}

Hasenfratz and Niedermayer \cite{hasenfratz} pointed out that optimal locality
properties of the perfect action $L_\Lambda $ are obtained for a
preferred finite value of $\kappa$.

\subsection{Evaluation of the perfect action,
  interpolation operator and the fluctuation field propagator}

To evaluate $L_\Lambda (\Phi )$ by perturbation expansion one proceeds
as follows \cite{mack}. One splits
\begin{eqnarray}
L(\varphi ) = L_0(\varphi ) + V(\varphi )
\end{eqnarray}

\begin{eqnarray}
L_0(\varphi ) = \frac 12\int \partial _\mu \varphi \partial _\mu
\varphi  = - \frac 12 (\varphi ,\triangle \varphi )
\end{eqnarray}
It is convenient to include mass terms in the interaction $V$.

Given the block spin $\Phi$, one determines a background field $\psi$
which minimizes $L_0$ subject to the constraint $C\psi = \Phi$.
Because $L_0$ is quadratic, $\psi$ is a linear function of $\Phi$.
\begin{eqnarray}
\psi (z)={\cal A} \Phi (z)=\int_{x \in \Lambda}{\cal A}(z,x)\Phi
(x),\ \ \ \ \mbox{with \thinspace \thinspace \thinspace \thinspace
\thinspace \thinspace \thinspace }C{\cal A}=1
\end{eqnarray}
and
\begin{eqnarray}
\int_{x \in \Lambda} = L_t L_s^3 \sum_{n \in {\menge N}^4},
\ \ \ \ \ \ \ \ \ x=({\bf n}L_s,n_4L_t).
\end{eqnarray}
In coordinate space
\begin{eqnarray}
v^{-1} {\cal A}=C^{\dagger}u^{-1},\ \ \ \ \ \ u=CvC^{\dagger},\ \ \ \ \ \ \ \ \
v=(-\Delta )^{-1}
\label{Aoper}
\end{eqnarray}

Following Gawedzki and Kupiainen \cite{gawedzki} one splits the field $\varphi
$ into
a low frequency part,
which is determined by the block spin $\Phi $ , and a high frequency
or fluctuation field $\zeta $ which has vanishing block average.
\begin{eqnarray}
\varphi ={\cal A}\Phi +\zeta, \,\,\,\,\,\,\,\,\ \ \ \ \ \ C\zeta =0
\label{phisplit}
\end{eqnarray}
Because of the extremality property of $\psi $, the kinetic energy
splits
\begin{eqnarray}
L_0(\varphi )=\frac 12({\cal A}\Phi ,-\Delta {\cal A}\Phi )+\frac
12(\zeta ,-\Delta \zeta )=\frac 12(\Phi ,u^{-1}\Phi )+\frac 12(\zeta
,-\Delta \zeta )
\end{eqnarray}
{}From this one sees that the free massless propagator for the block spin
$\Phi$ equals $u$. Since $C{\cal A}=1$, we have $\delta _\kappa
(C\varphi -\Phi )=\delta _\kappa (C\zeta ).$

The measure splits
\begin{eqnarray}
D\varphi \delta _\kappa (C\varphi -\Phi )\exp (-L_0(\varphi ))=D\zeta
\exp (-\frac 12(\Phi ,u^{-1}\Phi ))\delta _\kappa (C\zeta )\exp (-\frac
12(\zeta ,-\Delta \zeta ))
\end{eqnarray}
The $\zeta $-dependent factor is proportional to a Gaussian measure with
covariance (free propagator) $\Gamma $
\begin{eqnarray}
D\zeta \delta _\kappa (C\zeta )\exp (-\frac 12(\zeta ,-\Delta \zeta
))=Z^{-1}d\mu _\Gamma (\zeta )\ ,\ \ \ \ \ \ \ \ \ \ \Gamma =(-\Delta
+\kappa C^{\dagger}C)^{-1}
\end{eqnarray}
The limit $\kappa \rightarrow \infty $ can be taken if desired, and
results in \cite{gawedzki}
\begin{eqnarray}
\Gamma =v-{\cal A}u{\cal A}^{\dagger}=v-{\cal A}CvC^{\dagger}{\cal A}^{\dagger}
\quad \quad , \quad \quad C\Gamma=\Gamma C^\dagger = 0
\label{gamma}
\end{eqnarray}
The interpolation kernel ${\cal A}(z,x)$ of ${\cal A}$ has a Fourier
expansion like $C$
in eq. (\ref{ctilde}) with kernel
\begin{eqnarray}
\widetilde{{\cal A}}(l,p) = \tilde v(k)
\widetilde{C}^{*}(l,p)\mbox{}
\widetilde{u}\mbox{ }^{-1}(p)
\label{atilde}
\end{eqnarray}
\begin{eqnarray}
\widetilde{u}\mbox{ }(p)=\sum_l\tilde v(k) \mid \widetilde{C}(l,p)\mid
^2=\int_{x \in \Lambda}u(x)\exp (-ipx)
\label{utilde}
\end{eqnarray}
We use the abbreviation $\,\,\,k=p+l$ \thinspace \thinspace here and
throughout. We note that
\begin{eqnarray}
\tilde u(p)\sim \frac 1{p^2}
   \mbox{ \thinspace \thinspace \thinspace \thinspace
\thinspace \thinspace \thinspace \thinspace \thinspace \thinspace
\thinspace \thinspace \thinspace \thinspace \thinspace \thinspace
for\thinspace \thinspace \thinspace \thinspace \thinspace \thinspace
\thinspace \thinspace \thinspace \thinspace \thinspace \thinspace
\thinspace \thinspace \thinspace \thinspace \thinspace \thinspace
}p^2\rightarrow 0
\end{eqnarray}
because only the $l=0$ term in (\ref{utilde}) is singular at $p=0$.

The fluctuation field propagator $\Gamma (z,z^{^{\prime }})$ is
invariant under translations by lattice vectors in $\Lambda .$ It admits
therefore a Fourier expansion of the form
\begin{eqnarray}
\Gamma (z,z^{^{\prime }})=\sum_{l,l^{^{\prime }}}\int \frac{d^4p}{(2\pi
)^4} \exp (i(p+l)z-i(p+l^{^{\prime }})z^{^{\prime }})\widetilde{\Gamma
} (l,p,l^{^{\prime }})
\label{gammatilde}
\end{eqnarray}
Summations and integrations are as explained after (\ref{ctilde}).
All the summations over $l$ converge well.
For $\kappa
=\infty $ , eq. (\ref{gamma}) yields the explicit expression
\begin{eqnarray}
\widetilde{\Gamma }(l,p,l^{^{\prime }})=
\tilde v(k)\delta _{l,l^{^{\prime}}}
-\widetilde {{\cal  A}}(l,p)u(p)
\widetilde{{\cal A}}^{*}(l^{^{\prime }},p)
\end{eqnarray}
with $k = p+l$, and ${\cal A}$, $u$ from eqs. (\ref{atilde}) and
(\ref{utilde}).

Since $Z$ is constant, it follows that

\begin{eqnarray}
L_{eff}(\Phi )=\frac 12(\Phi ,u^{-1}\Phi )+V_{eff}(\Phi )+\mbox{const.}
\end{eqnarray}
\begin{eqnarray}
\exp (-V_{eff}(\Phi ))=\int d\mu _\Gamma (\zeta )\exp (-V_\Phi (\zeta
))\,\,\,\,\,\,\,\,\,\,\,\,\,\mbox{with\thinspace
}\,\,\,\,\,\,\,\,\,\,\,\, \,V_\Phi (\zeta )=V({\cal A}\Phi +\zeta ).
\label{veff}
\end{eqnarray}

Thus, V$_{eff}\ \ $is the free energy of a field theory with free
propagator $\Gamma $ and $\Phi $-dependent coupling constants. It admits
a standard perturbation expansion (cumulant expansion). We will write
down the leading terms for the $T > 0$ case
below. All this is well known.

\subsection{Scalar fields at finite temperature}

At finite temperature, we have periodicity in time with period $\beta $.
The extension $L_t$ of blocks in time direction must be chosen
commensurate with $\beta $.

A great simplification results if we chose $L_t=\beta $, so that only one
block fits in time direction. The lattice $\Lambda $ becomes a
3-dimensional lattice.

Our 3-dimensional fields still have the same dimension as the original
4-dimensional ones.
This could be remedied by a rescaling by $\beta^{\frac 12}$.
We prefer not to do so because eq.(\ref{leffsc}) below looks very natural.
Since we do not rescale the block spin $\Phi$ in position space the
appropriate 3-dimensional
integration measure contains a factor $\beta$ and the 3-dimensional
$\delta$-functions a corresponding factor $\beta^{-1}$, as follows.

\begin{eqnarray}
\int_{x \in \Lambda} = \beta L_s^3 \sum_{{\bf n} \in {\menge N}^3},
\ \ \ \ \ \ \ x=({\bf n}L_s,0).
\end{eqnarray}

In momentum space it is correspondingly.

To adjust to the finite temperature situation,
we need to periodize the $T=0$ quantities
in time in the manner described in section 2. Let us begin with the
fluctuation field propagator of eq.(\ref{gamma}).
We write $z=(\,{\bf z}, t)$.

\begin{eqnarray}
\Gamma _T(\,\,{\bf z}, t;{\bf z}^{^{\prime }},t^{^{\prime}})
&=& \sum_{n\in {\menge Z}}\Gamma (\,{\bf z}, t+n\beta ;{\bf
z}^{^{\prime}},t^{^{\prime }})
\end{eqnarray}

We insert the Fourier expansion for $\Gamma .$ Since $l_4, l_4^{^{
\prime }}\,\,\,\,\,\in 2\pi L_t^{-1}{\menge Z}\,\,\,
=2\pi \beta^{-1}{\menge Z}$, we have
$ l_4 t=l_4(t+n\beta )$ mod $2\pi $. Only a sum $\sum_n\exp
(ip_4(t+n\beta ))$ needs to be done. This sum was evaluated in section 2.
As a result

\begin{eqnarray}
\Gamma_T(\,\,{\bf z}, t;{\bf z}^{\prime },t^{\prime})
=\sum_{l,l^{\prime }}\frac 1{\beta} \int \frac{d^3{\bf p}}{\left( 2\pi \right)
^3}\exp(i {\bf k\cdot z}-i{\bf k}^{\prime }\cdot {\bf z}^{\prime
}+il_4t\,\,-\,\,il_4^{\prime }t^{\prime })\widetilde{\Gamma }(l,{\bf
p} ,0,l^{\prime })
\end{eqnarray}
where $\quad{\bf k}={\bf p} + {\bf l} \quad$ and
$\quad {\bf k^{\prime}}={\bf p} + {\bf l^{\prime}}$.

In other words, periodization has the effect of setting $p_4=0.$ This
reflects the fact that the lattice $\Lambda $ is only 3-dimensional.

Similarly one finds the interpolation and averaging kernels and the
block propagator.

\begin{eqnarray}
\widetilde{{\cal A}}_T(l,{\bf p})=\widetilde{{\cal A}}(l,{\bf
p,}0),\,\,\,\,\,\,\,\,\,\,\,\,\,\,\,\,
\,\,\,\,\,\,\,\,\,\,\,\,\,\,\,\,\,\,\,\,\,\,\,\,\,\,\,
\widetilde{C}_T(l,{\bf
p})=\widetilde{C}_s( {\bf l,p)}\delta _{l_4,0}
\end{eqnarray}
The block propagator in momentum space is $\tilde{u}({\bf p},0)$.
It is temperature dependent, because the range $l_4\in 2\pi L_t^{-1}
{\menge Z}$
of the $l_4$-summation depends on $\beta=L_t$.
For the same reason, the fluctuation field propagator and the
interpolation kernel are $T$-dependent.

It turns out that the temperature dependence of this block spin
propagator is a simple factor $\beta^{-1}$.
Therefore we agree to extract this factor, writing
\begin{eqnarray}
\tilde u_T (p) =
\tilde{u}({\bf p},0) = \beta^{-1} \tilde u_{FT\,\,}({\bf p})
\end{eqnarray}
with $T$-independent 'finite temperature' propagator
$\tilde u_{FT\,\,}({\bf p})$.
In coordinate space it is expressed in terms of the original bare
zero temperature propagator
\begin{eqnarray}
u_{FT}({\bf x}-{\bf y})
= \av_{{\bf z} \in [{\bf x}]} \av_{{\bf z^\prime} \in [{\bf y}]}
\int_0^\beta dt \, v_T({\bf z}-{\bf z^\prime},t)
= \av_{{\bf z} \in [{\bf x}]} \av_{{\bf z^\prime} \in [{\bf y}]}
\int_{-\infty}^{\infty} dt \, v_0({\bf z}-{\bf z^\prime},t)
\end{eqnarray}
$[{\bf x}]$ is the 3-dimensional cube of sidelength $L_s$ with central
point {\bf x}.

\subsection{Leading terms in the perturbation expansion of the perfect
action for $\varphi ^4$-theory at finite temperature}

The perfect lattice action at finite temperature equals

\begin{eqnarray}
L_{eff,T}(\Phi ) = \frac \beta 2 \left( \Phi ,u_{FT}^{-1}\Phi \right)
+V_{eff,T}(\Phi )
\label{leffsc}
\end{eqnarray}
\begin{eqnarray}
V_{eff,T}(\Phi ) = -\ln \left( \int d\mu _{\Gamma _T}\left( \zeta \right)
\exp \left( -V_{\Phi ,T}\left( \zeta \right) \right) \right)
\end{eqnarray}
with
\begin{eqnarray}
V_{\Phi,T}(\zeta )=V\left( {\cal A}_T \Phi +\zeta \right)
\end{eqnarray}
The effective interaction (including mass terms) is temperature
dependent because the fluctuation field propagator $\Gamma _T$ and the
interpolation kernel ${\cal A}_T$ are both temperature dependent.
This temperature dependence is weak and disappears in a zeroth order
local approximation.

Let us consider $\varphi ^4$-theory with bare mass $m_0$.

\begin{eqnarray}
L_0(\varphi )=\frac 12\int (\partial _\mu \varphi \partial _\mu \varphi
)\,\,dz
\end{eqnarray}

\begin{eqnarray}
\,\,\,V(\varphi )=\int \left( \frac 12m_0^2\varphi ^2+\frac \lambda
{4!}\varphi ^4\right) \,dz\,\,+\,\,\mbox{wave function renormalization term}
\end{eqnarray}
Inserting the split (\ref{phisplit}) of the field $\varphi $ , we obtain

\begin{eqnarray}
V_\Phi (\zeta )=U_{cl}(\Phi )+\sum_{n=1}^4\int dz\,\frac 1{n!}g_n(\Phi
,z)\,\,\zeta (z)^n
\end{eqnarray}
with

\begin{eqnarray}
g_1 &=& \frac \lambda {3!}\Psi (z)^3+m_0^2\Psi \left( z\right)
\nonumber \\
g_2 &=& \frac \lambda {2!}\Psi \left( z\right) ^2+ m_0^2
\nonumber \\
g_3 &=& \lambda \,\,\Psi \left( z\right)
\nonumber \\
g_4 &=& \,\lambda
\nonumber \\
U_{cl}\left( \Phi \right) &=& \int \left( \frac 12m_0^2\Psi \left( z\right)
^2+\frac \lambda {4!}\Psi \left( z\right) ^4\right) dz
\nonumber \\
\Psi \left( z\right) &=& \int_{x\in \Lambda }\,{\cal A\,}_T\left(
z,x\right) \Phi \left( x\right)
\end{eqnarray}
$V_{eff,T}$ can be calculated in a loop expansion.
When the starting point is a lattice theory, one can use Mayer
expansions instead.
They are convergent for weak coupling, and the asymptotic expansion of
individual terms in powers of the bare coupling constant contains
infinite sets of diagrams \cite{pordt}.

We write the perturbative expansion of $V_{eff,T}$ as

\begin{eqnarray}
V_{eff,T}\left( \Phi \right) =U_{cl}\left( \Phi \right) +\sum_{N\geq
1}V_{eff,T}^{(N)}\left( \Phi \right)
\end{eqnarray}
where $V_{eff,T}^{(N)}$ scales as $\gamma ^N$ when the fluctuation field
propagator $\Gamma _T\longrightarrow \gamma \,\,\Gamma_T $.

Let us compute $V_{eff,T}^{(1)}$.
Indicating factors $\Psi $ by dotted external lines
and a fluctuation field propagator by a solid line we have

%
\medskip
\centerline{\epsfxsize=350pt
\epsfbox{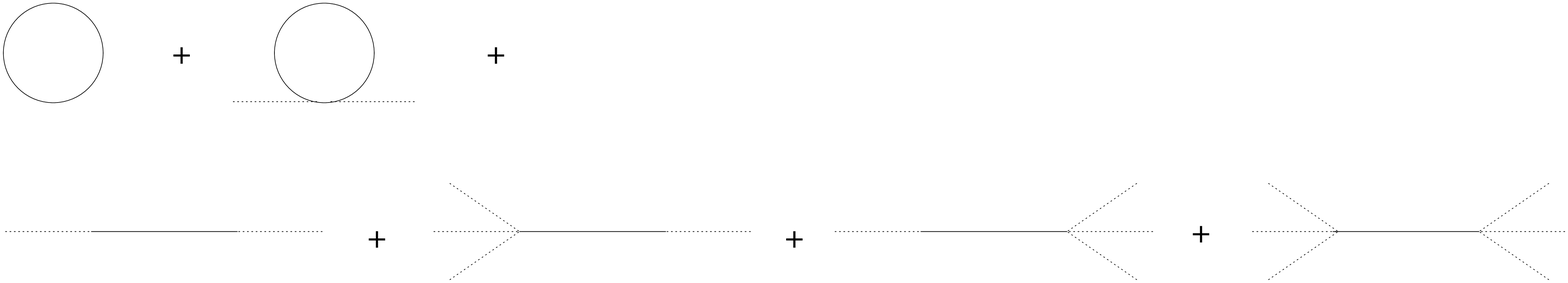}}
\refstepcounter{bild}
\label{sixpoint}
\biltit{sixpoint}{\ effective potential to first order in the
                                       fluctuation propagator}
\medskip
\begin{eqnarray}
V_{eff,T}^{(1)} &=&
 \frac 1 {2!} \int dz ( \frac \lambda {2!} \Psi(z) ^2 + m_o^2 )
\Gamma_T(z,z)
\nonumber \\
- \frac 1 {2!} \!\!\!\!\!\!\! & & \int dz_1 dz_2
( \frac \lambda {3!} \Psi(z_1) ^3 + m_o^2 \Psi(z_1) )
\Gamma_T(z_1,z_2)( \frac \lambda {3!} \Psi(z_2) ^3 + m_o^2 \Psi(z_2) )
\label{vefftsc}
\end{eqnarray}
The fluctuation field propagator $\Gamma _T(z,z^{^{\prime }})$, decays
exponentially with distance $\left| z-z^{^{\prime }}\right| $ with decay
length one block lattice spacing. That is, it decays with $\left| {\bf
z}- {\bf z}^{^{\prime }}\right| $ with decay length $L_s$. Similarly,
$\,{\cal A}_T (z,x)$ decays exponentially in $\left| z-x\right| $ with
decay length $L_s$. As a result, each term in $V_{eff,T}$ is local in
$\Phi $ modulo exponential tails with decay length of one lattice
spacing.

By a process of partial integration (or rather summation) one can
exhibit each term in $V_{eff,T}$ as a sum of local terms and small
remainders which represent the exponential tails (see appendix \ref{local}).
The terms with
coefficients of dimensions up to (mass$)^{-2}$ are

\begin{eqnarray}
V_{eff,T}\left( \Phi \right) =\int_x\left\{ \frac 12m^2\Phi ^2+\delta
_z\left( \nabla _\mu \Phi \right) ^2+\frac{\lambda _r}{4!}\Phi
^4+\frac{ \lambda _6}{6!}\Phi ^6+
\tilde \gamma \left( \nabla _\mu \Phi \right) ^2\Phi ^2+ \cdots \right\}
\end{eqnarray}
Since $\tilde u_{FT}^{-1}(p)\sim p^2$ as $p\rightarrow 0$, we may write the
expansion equally well in the form
\begin{eqnarray}
V_{eff,T}\left( \Phi \right) &=& \int_x\left\{ \frac 12m^2\Phi
^2+\frac{\lambda _r}{4!}\Phi ^4+\frac{\lambda _6}{6!}\Phi
^6 \right\}
\nonumber \\
&+& \frac \beta 2 \int_x\int_y\Phi
(x)u_{FT}^{-1}(x,y)\Phi (y)\left[ 2 \delta_z+ \gamma \Phi(x)^2\right] +\cdots
\label{scvefft}
\end{eqnarray}
All the coefficients are $T$-dependent and finite.

For computer simulations it is appropriate to consider $V_{eff,T}$ as a
function of $\Phi (x)$. For analytical computations it is most
convenient to regard it as a function of $\Psi (z)$.

\subsection{Note on the cancellation of temperature dependent
UV-divergent diagrams}

The UV-convergence of a quantum field theory concerns its local behaviour
and is therefore not temperature dependent. If the proper choice of
counter terms makes the theory finite at zero temperature, then also at
finite temperature. The cancellation occurs order by order in
perturbation theory. This is well known.

Problems can occur when one sums selected classes of diagrams. We do not
propose to do so when deriving the perfect action, but stay strictly
within the realm of standard perturbation theory. So there can be no
problem.

It is nevertheless appropriate to point out that there do exist
individual diagrams with temperature dependent UV-divergent pieces. They
cancel. We give an example.

Let us split the (fluctuation field) propagator $\Gamma $ into a static
part (heavy line) $\Gamma _{stat}$ and a non static part $\Gamma
_{ns}$ (wavy line). The non static part represents random walks which
wind several times around the tube; it is not singular at coinciding
arguments : $\Gamma _{ns}(z,z^{^{\prime }})<\infty $ . But it is not
zero and is $T$-dependent. As a result, the diagrams

%
\medskip
\centerline{\epsfxsize=250pt
\epsfbox{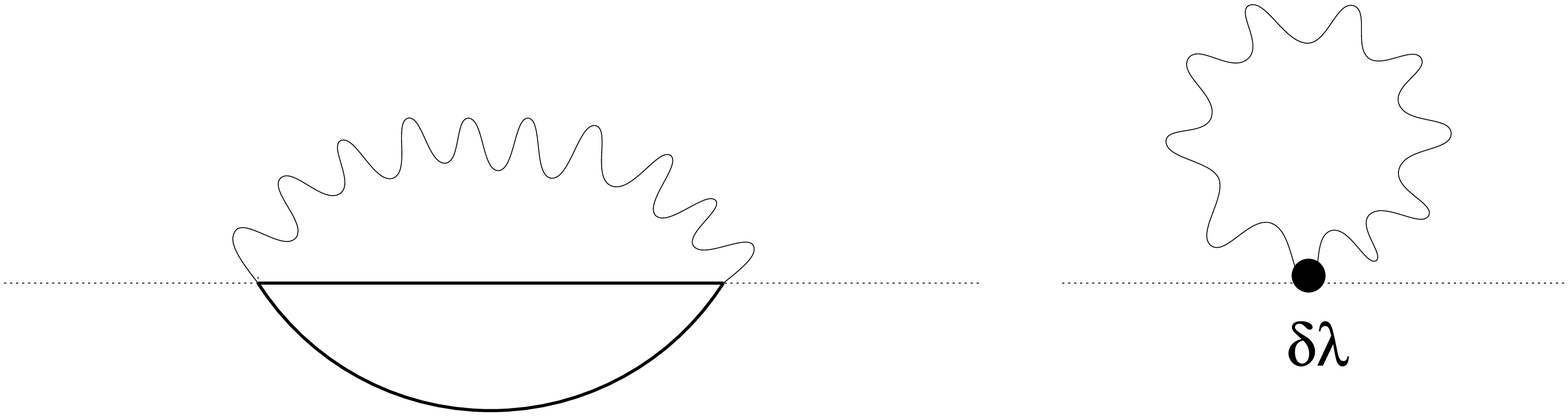}}
\refstepcounter{bild}
\label{divergenz}
\biltit{divergenz}{\ temperature dependent divergent diagrams}
\medskip
\noindent%
are both logarithmically UV-divergent, with a temperature dependent
coefficient $\Gamma _{ns}(z,z^{\prime })$ . They must cancel. $\delta
\lambda $ is the logarithmically divergent coupling constant counter
term to second order in $\lambda $ .

\section{The Feynman-Bogoliubov Method}
\label{secfeyman}

To obtain information on the nature of a phase transition at finite
temperature, one wants to compute the effective potential, i.e. the free
energy as a function of the magnetization.

\begin{eqnarray}
M=\int dz\varphi (z)=\int_x\Phi (x)
\end{eqnarray}

Alternatively, one may compute the constraint effective potential which
gives the probability distribution of $M$.

To do so we should apply nonperturbative methods to solve the lattice
theory.
One of these methods consists in the solution of gap equations. It is very
old and known as the Feynman-Bogoliubov method \cite{feybo}.

The gap equations have ''perturbative'' solutions which come from
summations of superdaisy diagrams. In principle they might also have other
solutions. We wish to examine this method in order to see what will be
the effect of terms like $\left( \nabla _\mu \Phi \right) ^2\Phi ^2$
etc. in the perfect lattice action.

Let us make it clear that it is not sufficient to find the solutions of
these gap equations. In order to justify the perturbative calculation of
the perfect action it will be necessary to investigate also the
stability properties of the solutions of the gap equations against small
perturbations of the lattice action. We will comment on this, but a
thorough treatment of this question is beyond the scope of this paper.

Given an action $S(\Phi )$ of some Euclidean field theory, one seeks an
optimal quadratic approximation $S_{free}(\Phi )$ around which to expand

\begin{eqnarray}
S_{free}\left( \Phi \right) =\frac 12\int_x\int_y\Phi (x)J(x,y)\Phi (y)
\end{eqnarray}
By Peierls inequality \cite{feybo}, the partition functions obey the inequality

\begin{eqnarray}
\ln \left( Z\right) \geq \,\ln (Z_{free})\,\,\,\,-\,\,\left\langle
S-S_{free}\right\rangle _{free}
\label{peierls}
\end{eqnarray}
for any choice of the $S_{free}.\,$Herein, $<>_{free}\,$is the
expectation value in the theory with action $S_{free}.$ The right hand
side is the first order approximation to the left hand side in the
perturbative expansion around $S_{free}$. The optimal choice of
$S_{free}$ is that which makes the right hand side of (\ref{peierls})
 maximal. In
other words, it makes the first order approximation as good as possible.
It is asserted that there exists always a unique optimal choice of
$S_{free}$. It is not asserted that the optimal choice is necessarily a
good one. The optimal $J$ is determined by the extreme value condition

\begin{eqnarray}
\left\langle \frac{\delta ^2S}{\delta \Phi \left( x\right) \delta \Phi
(y)} \right\rangle _{free}=\,\,\,J(x,y)
\label{gap}
\end{eqnarray}
This is equivalent to the condition that the right hand side of
(\ref{peierls}) is maximal, viz

\begin{eqnarray}
\frac \delta {\delta J}\left[ \ln (Z_{free})-\left\langle
S-S_{free}\right\rangle _{free}\right] =0
\end{eqnarray}
When applied to the standard $\Phi ^4$ action, this produces the gap
equation whose perturbative solution is the sum of superdaisy diagrams
(see below).

Let us consider the gap equation which results from the more complicated
lattice action.

\begin{eqnarray}
S(\Phi )=\frac {\beta}2(\Phi ,u_{FT}^{-1}\Phi )+V_{eff,T}\left( \Phi \right)
\label{sphi}
\end{eqnarray}
with $V_{eff,T}$ from equation (\ref{scvefft}).

We obtain
\begin{eqnarray}
\frac{\delta ^2S(\Phi )}{\delta \Phi \left( x\right) \delta \Phi (y)}
 &=& \beta u_{FT}^{-1}(x,y) \left[ 1+2 \delta_z+ 3\gamma \Phi(x)^2 \right]
\nonumber \\
&+& \beta \delta (x-y) \,3\, \gamma\, \Phi(x) (u_{FT}^{-1}\Phi)(x)
\nonumber \\
&+& \delta (x-y) \left(m^2 + \frac{\lambda _r}{2!} \Phi (x)^2
+ \frac{\lambda_6}{4!} \Phi (x)^4  \right)
\end{eqnarray}
The expectation value in the theory with action $S_{free}$ $=\int \frac
12\Phi J\Phi $ is
\begin{eqnarray}
\left\langle \frac{\delta ^2S(\Phi )}{\delta \Phi \left( x\right) \delta
\Phi (y)}\right\rangle _{free} &=&
\beta u_{FT}^{-1}(x,y) \left[ 1 +2 \delta_z+ 3 \gamma J^{-1}(0) \right]
\nonumber \\
&+& \beta \delta (x-y) 3\, \gamma\, (u_{FT}^{-1}J^{-1})(0)
\nonumber \\
&+& \delta (x-y) \left( m^2+\frac {\lambda_r} 2
J^{-1}(0) + \frac{\lambda _6}{2^2 2!}J^{-1}(0)^2  \right)
\end{eqnarray}
where $J^{-1}(0) = J^{-1}(x,x) $ is
independent of $x$ by translation invariance, assuming that we seek a
translation invariant solution. Other solutions could be of interest.

The gap equation (\ref{gap}) can be solved by the Ansatz

\begin{eqnarray}
J(x,y)=Au_{FT}^{-1}(x,y)+B\delta (x-y)
\end{eqnarray}

Inserting the Ansatz results in two transcendental equations for $A,B$
whose solutions depend on the coefficients $\beta,\delta_z,\gamma ,\lambda
_i.$

Basically, the inclusion of the $\left( \nabla _\mu \Phi \right) ^2\Phi
^2$ -term results in a system of two equations for mass and wave
functions renormalization. In standard $\Phi ^4$-theory there is only
one equation for the mass.

\subsection{Comparison with other approaches}

The standard method of dealing with the finite temperature phase
transition of $\Phi ^4$-theory is described by Kapusta \cite{kapusta}. It
consists in the summation of ring or daisy diagrams which contribute to
the self energy part $\Pi$ of the physical finite temperature
propagator.

It would seem natural to improve this by solving the gap equations
instead. This would replace the sum over daisy diagrams by a sum over
superdaisy diagrams. The temperature dependent mass $m^2\left( T\right)
$ would be determined in a self consistent way. But there is a problem
with this: There arises a temperature dependent logarithmically
divergent contribution to the self energy which cannot be cancelled by a
temperature-independent bare mass term. The approach outlined in this
paper avoids this problem, because UV-divergences are cancelled before
one derives the gap equation. We see here the advantage of making a
clean separation between UV- and IR-problems.

Let us explicitely show how the problem arises in a conventional
superdaisy approach without a lattice as an UV-cutoff.

We use the standard $\Phi ^4$-action in place of (\ref{sphi}) and make the
usual Ansatz for the solution of the gap equation.

\begin{eqnarray}
J(x,y)=-\Delta \delta (x-y)+m^2(T)\delta (x-y)
\end{eqnarray}
We obtain the gap equation (\ref{gap}) as before with $\delta _z=\gamma
=\lambda_6=0$, $u^{-1}=-\Delta $. It is solved if

\begin{eqnarray}
m^2(T)=m_0^2+ \frac {\lambda_r} 2 J^{-1}(0)
\end{eqnarray}

\begin{eqnarray}
J^{-1}(0) = T\sum_n\int \frac{d{\bf p}}{(2\pi )^3}\left[
\omega _n^2+{\bf p}^2+m^2(T)\right]^{-1}
\end{eqnarray}
The integral defining $J^{-1}(0) $ is evaluated in the
standard fashion, resulting in

\begin{eqnarray}
J^{-1}(0) =\Pi _{vac}+\Pi _{mat}
\end{eqnarray}

\begin{eqnarray}
\Pi _{vac}=\int \frac{d^4p}{(2\pi )^4}\left[ p^2+m^2(T)\right]
^{-1}=\frac 1{16\pi ^2}\left[ \Lambda ^2-m^2(T)\ln \frac{\Lambda
^2}{m^2(T)}\right]
\label{pivac}
\end{eqnarray}

\begin{eqnarray}
\Pi _{mat}=\int \frac{d^3{\bf p}}{(2\pi )^3}\frac 1\omega \left[ \exp
(\beta \omega )-1\right]
^{-1}\,\,\,\,\,\,\,;\,\,\,\,\,\,\,\,\,\,\,\,\,\,\,\,\omega =({\bf
p}^2+m^2(T))^{\frac 12}
\end{eqnarray}
The integrals are UV-divergent, therefore a regulator $\Lambda$ is
introduced.
It is supposed to be taken to infinity after cancellation against
counterterms.
We see that without a lattice a temperature dependent counterterm is needed
to cancel the logarithmic UV-divergence in (\ref{pivac}).
This is not acceptable.

Conclusion: The Feynman-Bogoliubov method is designed to deal with
infrared problems. It should only be applied after a correct
cancellation of UV-divergences has been achieved through the computation
of an effective field theory, e.g. on a lattice, or with another low
UV-cutoff.

We wish to compare our proposal with another approach. It was proposed
to derive first a 3-dimensional continuum theory. When no
UV-regularization is introduced, then this 3-dimensional theory is
superrenormalizable but not finite. It has linear self energy
divergences. These divergences are cancelled by divergent terms in the
3-dimensional action. Again the problem remains that the UV-divergences
have not been completely cancelled before nonperturbative methods are
applied to deal with infrared aspects.
In addition, the 3-dimensional continuum theory is complicated because
of nonlocalities with decay length $\frac \beta {2\pi}$
which appear as soon as one goes beyond the leading order in the
perturbative expansion of the 3-dimensional action
(see appendix \ref{dimred}).
The use of a momentum expansion to deal with these nonlocalities would
aggravate the UV-problems.

The 'average action' approach of Wetterich \cite{wett} is in the same spirit
as ours. The difference is that we use a lattice. The lattice has
the advantage that one can do computer simulations.
Wetterichs work suggests moreover, that one should perform
simplifications only on the 1-particle irreducible parts of the perfect
action (i.e. those not held together by a single $\Gamma$-propagator).
It has been known for a long time that the 1-particle reducible second
term $\Psi^3 \Gamma_T \Psi^3$ in eq.(\ref{vefftsc}) is important
inspite of looking irrelevant \cite{gawedzki}.

\section{Perfect lattice action for Maxwell theory }
\label{secmaxwell}

We wish to extend the consideration of the previous sections to scalar
electrodynamics. As a preparation for this we consider
the free abelian gauge field.

In this section we recall the Ba{\l}aban-Jaffe block spin transformation
for the free abelian gauge field at zero temperature.

\subsection{Averaging operators for abelian gauge fields}

Given a vector potential $a(z)=a_\mu (z)dz_\mu $ on the continuum, we
define a block spin ${\bf A}$ on the block lattice. ${\bf A}$ lives on
links $b$ of the block lattice. We use the alternative notation

\begin{eqnarray}
{\bf A}[b]={\bf A}_\mu (x)
\end{eqnarray}
when $b$ is the link emanating from $x$ in $\mu $-direction.

We will not distinguish in notation between the averaging operator C for
gauge fields, and for scalars (gauge transformations). Which one is meant
will be clear from the context. Thus ${\bf A}=Ca$.

Given a link $b$ from $x$ in $\mu $-direction, and a point $z\in x$, let
${\cal C}_{z,\mu }$ be the straight path of length one block lattice spacing
in $\mu $-direction starting from $z$. If $\mu =4$,
${\cal C}_{z,\mu }$ connects $z$ with
$ z+L_te_4,$ and if $\mu \neq 4$ it connects $z$ with $z+L_se_\mu
$ ($e_\mu =$ unit vector in $\mu$-direction). The blocking procedure is
defined by

\begin{eqnarray}
{\bf A} \left[ b\right] =\av_{z\in x}\,a\left[{\cal C}_{z,\mu
}\right]
\end{eqnarray}

\begin{eqnarray}
a \left[{\cal C}_{z,\mu }\right] =
\int_{{\cal C}_{z,\mu }}dw_\nu a_\nu (w)
\end{eqnarray}
This blocking procedure is covariant under gauge transformations in the
following sense. If

\begin{eqnarray}
a_\mu (z)\rightarrow a_\mu (z)-\partial _\mu \lambda (z)\equiv a_\mu
^\lambda (z)
\end{eqnarray}
then
\begin{eqnarray}
{\bf A}_\mu (x)\rightarrow {\bf A}_\mu (x)-\nabla _\mu \Lambda (x)
\end{eqnarray}

\begin{eqnarray}
\Lambda =C\,\,\lambda
\end{eqnarray}
i.e.
\begin{eqnarray}
\Lambda (x)=\av_{z\in x}\,\lambda (z)
\end{eqnarray}
We use the notation $a^\lambda $ for the gauge transform of $a$ , etc.
In this notation, the covariance property reads

\begin{eqnarray}
C\,a^\lambda =(C\,a)^{C\,\,\lambda }
\label{gtrafo}
\end{eqnarray}
We will need the Fourier transform of the $C$-kernel.
Let us write

\begin{eqnarray}
{\bf A}_\mu (x)=\int C_{\mu \nu }(x,z)\,a_\nu (z)
\end{eqnarray}
The kernel $C_{\mu \nu }$ will have a Fourier expansion just like
the scalar kernel. Explicitly
\begin{eqnarray}
\widetilde{C}_{\mu \nu }(l,p)=\delta _{\mu \nu }\exp (-i\,\frac {k_\nu L_\nu}
2 )\,\,\frac 2{L_\nu \,k_\nu }\,\,\sin \frac{k_\nu L_\nu}
 2 \,\,\, \tilde C(l,p)
\end{eqnarray}
where $\tilde C(l,p)$ is the scalar kernel (\ref{ctilde}), and
no sum over $\nu$ is implied. $L_\nu =\,L_t$ if $\nu =4$ and $L_\nu
=\,L_s$ otherwise; $k_\nu = p_\nu + l_\nu$ as usual.

\subsection{Perfect lattice action for the free electromagnetic
field}

Using the symbol $\partial $ for the exterior derivative, $\partial
a=\frac 12\left( \partial _\mu a_\nu -\partial _\nu a_\mu \right)
dz_\mu \wedge dz_\nu $, the Maxwell action of the
electromagnetic field can be written as

\begin{eqnarray}
S_M(a)=\frac 12<\partial a,\partial a>
\end{eqnarray}

The perfect lattice action associated with this is defined by a formula
analogous to eq. (\ref{veff}) for the scalar field. There is one difference,
however. In order to give meaning to the functional integral, some
amount of gauge fixing is necessary.

We wish to obtain a gauge covariant perfect lattice action. Therefore we
wish to retain the freedom of gauge transformations on the lattice, i.e.
one gauge degree of freedom per block. Therefore global gauge fixing is
not appropriate. Instead, gauge fixing is only used locally within each
block.

Consider gauge transformations

\begin{eqnarray}
a_\mu \rightarrow a_\mu -\partial _\mu \,\lambda
\,\,\,\,\,\,\,\,\,\,\,\,\,\,\,\,\mbox{with}\,\,\,\,\,\,\,\,\,\,\,\,\,\,\,\,
\,C\,\lambda =0
\end{eqnarray}

By eq. (\ref{gtrafo}), they leave the block gauge field invariant. These are
the
gauge degrees of freedom which will be eliminated by fixing the gauge
within a block.
There remains one gauge degree of freedom per block $\Lambda
(x)=C\,\,\lambda (x)$ which is not affected by the fixing. It extends to
a global gauge transformation per block, $\lambda (z)=\Lambda (x)$ for
$z\in x.$

To begin with, a block axial or radial gauge is used. For the purpose of
perturbative calculations, a transformation to a block Landau gauge is
carried out later.

Suppose for a moment that we start from a theory on a fine lattice
$\Lambda _{fine}$ instead of the continuum. $a$ will then live on links
$l$ of $ \Lambda _{fine}.$ We describe the block axial gauge for this
situation first. The formal continuum limit will be obvious.

For every block $y\in \Lambda $ , a maximal tree $Ax(y)$ is selected,
for instance the comb within each block shown in the
figure below.

%
\medskip
\centerline{\epsfxsize=100pt
\epsfbox{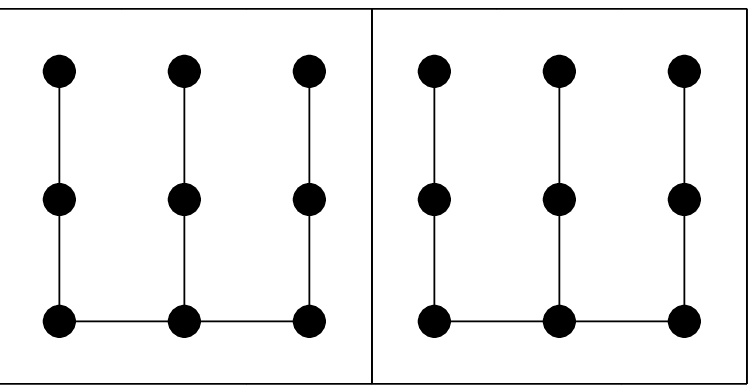}}
\refstepcounter{bild}
\label{comb}
\biltit{comb}{\ maximal trees within 2-dimensional blocks}
\medskip

$Ax(y)$ consist of links $l\in \Lambda _{fine}$. In the block axial
gauge $ a\left[l\right] =0$ for all $\,\,l\in Ax(y),\,\,\,\,y\in
\Lambda .$

One defines appropriate $\delta$-functions which are inserted in the
functional integral

\begin{eqnarray}
\delta _{Ax(y)}(a)=\prod_{l\in Ax(y)}\delta \left( a\left[ l\right]
\right)
\end{eqnarray}

\begin{eqnarray}
\delta _{Ax}(a)=\prod_{y\in \Lambda }\delta _{Ax(y)}(a)
\end{eqnarray}
The perfect lattice action $S_{eff}({\bf A})$ is defined up to an additive
constant by

\begin{eqnarray}
\exp (-S_{eff}({\bf A}))=\int Da\,\delta _{Ax}(a)\,\delta
\left( C\,a-{\bf A}\right)\,\exp (-S_M(a))
\label{seffaax}
\end{eqnarray}
Instead of the block axial gauge, the block radial gauge can be used. In
the continuum limit it is

\begin{eqnarray}
\sum_{\mu =1}^4(z-x)_\mu \,\,a_\mu
(z)\,\,=\,\,0\,\,\,\,\,\,\,\,\,\,\,\,\,\,\,\,\,\,\,\,\,\,\,\,\,\,\,\,\,\,\,
\,\,\,\,\,\,\,\,\,\,\,\,{\rm for}\,\,\,\,\,\,\,\,z\in x \quad .
\end{eqnarray}

\subsubsection{Transformation to the block Landau or $\alpha $-gauge}

For perturbative calculations, the block axial gauge is inappropriate
because of the bad ultraviolet behaviour of the fluctuation field
propagator in this gauge. Therefore the Faddeev-Popov method is used to
convert to a block Landau or $\alpha $-gauge.

$S_M$ is invariant under arbitrary gauge transformations while $\delta
\left( C\,a-A\right) $ is only invariant under those gauge transformations $
a\rightarrow a^\lambda =a-\partial \,\,\lambda $ which obey

\begin{eqnarray}
C\,\lambda =\,0
\label{blockgauge}
\end{eqnarray}
$C$ is the averaging of a scalar function over blocks. We wish to
eliminate the freedom of performing gauge transformations which obey
(\ref{blockgauge}) by gauge fixing.

Recalling that $\partial$ stands for the exterior derivative of a p-form,
$\partial^{\dagger}$ is the coderivative.
For example $\partial^{\dagger} a = \partial_\mu a_\mu$ and
$\partial^{\dagger} \lambda =0$ if $\lambda$ is a 0-form.
In this language the Laplacian is given by $\partial\partial^{\dagger}
+\partial^{\dagger}\partial$.

To convert to the block $\alpha $-gauge, one multiplies and divides
expression (\ref{seffaax}) by

\begin{eqnarray}
V (a) = \int D\lambda \,\,\delta \left( C \lambda \right) \exp \left\{ -\frac
1{2\alpha }<\partial ^{\dagger}a^\lambda ,\,\partial ^{\dagger}a^\lambda
>\right\}
\end{eqnarray}
After interchanging the order of integration and performing a gauge
transformation $a \rightarrow a^{-\lambda }$ in the $a$-integral

\begin{eqnarray}
\exp (-S_{eff}({\bf A}))&=&\int Da\,\frac 1 {V(a)} \,\delta \left(
C\,a-{\bf A}\right) \exp
(-S_M(a)-\frac 1{2\alpha }<\partial ^{\dagger}a ,\,\partial
^{\dagger}a >)
\nonumber \\
& & \int D\lambda \delta (C\lambda )\delta _{Ax}(a^{-\lambda })
\label{seffal}
\end{eqnarray}
The second factor is 1 because $\lambda $ is fixed up to a global
transformation by the condition
$\delta(a^{-\lambda}[l])$
for every link
$l\in Ax(y)$ in the maximal tree of block $y$, and the remaining global
transformation is absorbed by the $\delta$-function $\delta (C\lambda
).$

It remains to evaluate

\begin{eqnarray}
V(a)=\int D\lambda \,\,\delta \left( C\lambda \right)   \exp \left\{ -\frac
1{2\alpha }<\partial ^{\dagger}a^\lambda ,\,\partial ^{\dagger
}a^\lambda >\right\}
\end{eqnarray}
\begin{eqnarray}
=\int D\lambda \delta (C\lambda )\exp
\left\{- \frac 1 {2\alpha}
<(\Delta^{-1}\partial^{\dagger}a-\lambda),\Delta^{\dagger} \Delta
(\Delta^{-1}\partial^{\dagger}a -\lambda)> \right\}
\end{eqnarray}
Thinking of the $\delta $-function as a limit of a Gaussian, one sees
that this is a Gaussian integral.
But since the $\delta$-function restricts the range of the integration the
result is not just a constant.
To show this we introduce the projector $R$ onto those functions
$\lambda$ which satisfy the constraint $C\lambda=0$.
$R$ can be chosen orthogonal with respect to the scalar product
$<\cdot\,\,\,,\Delta^{\dagger} \Delta \,\,\,\cdot>$.
The integral is split into two parts, the integration over $(1-R)\lambda$
is restricted by the $\delta$-function to $(1-R)\lambda=0$ whereas the
integration over $R\lambda$ has no further restriction.
Since $R$ is an orthogonal projector there exists no mixed term and
the integral is a pure Gaussian.

\begin{eqnarray}
V(a)=\mbox{const}\,\,\exp \left\{ -\frac 1{2\alpha }<\partial
^{\dagger}a,\Delta (1-R)\Delta^{-1}\partial ^{\dagger}a > \right\}
\end{eqnarray}
Inserting this into eq.(\ref{seffal}) we obtain the final result

\begin{eqnarray}
\exp (-S_{eff}({\bf A}))=\int Da\,\,\delta \left( C\,a-{\bf A}\right) \exp
(-S_M(a)-\frac 1{2\alpha }<\partial ^{\dagger}a,\Delta R
\Delta^{-1}\partial ^{\dagger}a>)
\end{eqnarray}

The projector $R$ ensures that only the gauge degrees of freedom
$\lambda $ with $C\,\lambda =0$ -- i.e. vanishing block averages of
$\lambda $ -- are fixed. There remains the freedom of performing global
gauge transformations on each block. This freedom is reflected in the
fact that $\exp (-S_{eff}(A)) $ is invariant under gauge transformations
${\bf A}\rightarrow \,{\bf A}^\Lambda ={\bf A}-\partial \,\Lambda $,
where $\Lambda $ is a scalar function on the lattice.

\subsubsection{Explicit formula for $R$}

We recall from section 3 that the fluctuation field $\zeta $ associated
with a scalar field $\phi $ satisfies $C\,\zeta =0$ and can be obtained
by applying a projector

\begin{eqnarray}
\zeta =(1-{\cal A\,\,}C)\,\,\varphi
\end{eqnarray}
${\cal A\,\,}C$ is an orthogonal projector with respect to the scalar
product which is furnished by the kinetic term for $\varphi $.

Here we look for an orthogonal projector with respect to the scalar product
employing the kinetic term for
$ \lambda $ which is  $\Delta ^2$ in place of  $-\Delta $ . Therefore we
may write

\begin{eqnarray}
R=1-{\cal A}^{(\lambda )}\,\,\,C
\end{eqnarray}
where $C$ is the same averaging kernel
for scalars as before with its Fourier
transform given by eq.(\ref{ctilde}), while
${\cal A}^{(\lambda)}$ is chosen to satisfy
\begin{eqnarray}
v^{-1}{\cal A}^{(\lambda)}=C^{\dagger}u^{-1}; \,\,\,\,\,\,\,\, u=CvC^{\dagger};
\,\,\,\,\,\, v^{-1}=\Delta^2
\end{eqnarray}
in complete analogy to eq.(\ref{Aoper}).

The Fourier transform is
\begin{eqnarray}
\widetilde{{\cal A}}^{(\lambda)}(l,p)=\tilde v(k)
\widetilde{C}^*(l,p)\,\,\tilde u^\lambda(p)^{-1}
\end{eqnarray}
with \,\,\,\,\,\,\,\,\,\ $k=l+p ,\,\,\,\,\,\,\,\,\,\,\,\,\,\,\,
\tilde v(k)=\frac 1 {k^4}$ \,\,\,\,\,\,\,\,\, and
\begin{eqnarray}
\tilde u^{\lambda }(p)=
 \sum_l \tilde v(k) \left|\widetilde{ C}(l,p)\right| ^2
\end{eqnarray}

\subsection{Interpolation kernel ${\cal A}$ and evaluation of the
perfect lattice Maxwell action}

The perfect lattice action is determined in essentially the same way as
for the scalar field.

Given the block gauge field ${\bf A}$, one seeks that gauge field
$\hat a$ which
minimizes $S_M(a)$ $+\frac 1{2\alpha }<\partial ^{\dagger}a,\Delta R
\Delta^{-1} \,\partial^{\dagger}a>$ subject to the constraint
$C\,\hat a = {\bf A}$.
Because the action is quadratic, $\hat a$ is a linear function of ${\bf A}$

\begin{eqnarray}
\hat a={\cal
A\,}{\bf A}\,\,\,\,\,\,\,\,\,\,\,\,\,\,\,\,\,\,\,\,\,\,\,\,\,\,\,\,\,\,\,
\,\,\,\,\,\,i.e.\,\,\,\,\,\,\,\,\,\,\,\,\,\,\,\,\,\,\,\,\hat a_{\mu}
(z)=\int_{links\,\,b\in \Lambda }\,{\cal A\ }_\mu(z,b)\,\,{\bf A}[b]
\end{eqnarray}

Ba{\l}aban and Jaffe show that $\hat a$ -- and therefore ${\cal A\ }$-- are
independent of $\alpha $ and satisfy the block Landau gauge condition

\begin{eqnarray}
{\cal R} \,\partial^{\dagger}\,\hat a=0\,\,\,\,\,\,\,\,\,
\,\,\,\,\,\,\,\,\,\,\,\,\,\,\,\,\,\,\,
\,i.e.\,\,\,\,\,\,\,\,\,\,\,\,\,\,\,\,\,\,\,\,\,\,\,
{\cal R} \,\partial\,^{\dagger}\,{\cal A\ }=0
\label{proj}
\end{eqnarray}
where ${\cal R}$ is a shorthand notation for $\Delta R \Delta^{-1}$.
The proof is based on constructing a gauge transformation $\lambda $
with $ C\,\lambda \,=0$ which ensures (\ref{proj}). The equation

\begin{eqnarray}
\Delta \,\lambda ={\cal R} \,\partial \,^{\dagger}\,a
\end{eqnarray}
has a unique solution in the subspace $C\,\lambda \,=0$ because ${\cal R}$
is the projector onto those
configurations $\Delta \lambda$ which satisfy $C\,\lambda \,=0.$

The kernel ${\cal A}_\mu (z,b)$ has exponential decay in the distance of
$z$ from the blocks $(x,y)=b \quad$ \cite{balaban}.

The effective action is obtained by shifting the field $a=\,{\cal A\
} {\bf A}+a^{^{\prime }}$ where $a^{^{\prime }}$ is called the fluctuation
field. The integral over $a^{^{\prime }}$ produces merely a constant and
one obtains
\begin{eqnarray}
\exp (-S_{eff}({\bf A}))=\exp \left( -\frac 12<\partial {\cal A}{\bf A},
\partial
{\cal A} {\bf A}>-\frac 1{2\alpha }<\partial ^{\dagger}{\cal A}{\bf A},
{\cal R} \partial^{\dagger}{\cal A}{\bf A}>\right)
\end{eqnarray}

\begin{eqnarray}
=\exp \left( -\frac 12<\partial {\cal A} {\bf A},\partial {\cal A} {\bf A}>
\right)
\end{eqnarray}
independent of $\alpha$
because of eq. (\ref{proj}). Clearly $S_{eff}({\bf A})$ is a bilinear function
of ${\bf A}$. One may therefore define an operator $\Delta _1$ acting
on block gauge field configurations $ {\bf A} $, such that

\begin{eqnarray}
S_{eff}( {\bf A})=\frac 12< {\bf A},\Delta _1 {\bf A}>
\end{eqnarray}
Explicitly

\begin{eqnarray}
\Delta _1={\cal A}^{\dagger}{\cal \,\,}\partial \,^{\dagger
}\,\partial \,\,{\cal A}
\end{eqnarray}

\subsection{Fluctuation field propagator}

The probability distribution of the fluctuation field is given up to a
normalization factor by $(\kappa =\infty )$
\begin{eqnarray}
\delta _\kappa (C\,a^{^{\prime }})\,\exp \left\{ -\frac
12<\partial a^{^{\prime }},\partial a^{^{\prime }}>-\frac
1{2\alpha }<\partial^{\dagger} a^{^{\prime }},{\cal R}
\,\partial^{\dagger}
a^{^{\prime}}>\,\right\}
\end{eqnarray}
Thinking again of the $\delta $-function as a limit of a Gaussian
$\delta _\kappa$

\begin{eqnarray}
\delta _\kappa (C a^{\prime})= \left( \frac{2\pi }\kappa \right) ^{-\frac
12}\,\,\exp \left( -\frac \kappa 2<C a^{\prime},C a^{\prime}>\right)
\end{eqnarray}
we see that the probability distribution is given by a Gaussian measure
with covariance

\begin{eqnarray}
\Gamma _\alpha =\lim _{\kappa \rightarrow \infty }\,\,\Gamma _{\kappa
,\alpha }
\end{eqnarray}

\begin{eqnarray}
\,\Gamma _{\kappa ,\alpha }=\left( \partial \,^{\dagger}\partial +\frac
1\alpha \partial \,\,{\cal R}\,\,\partial ^{\dagger }+\kappa \,\,C\,^{\dagger}
C \right) ^{-1}
\end{eqnarray}
$C$ is again the Ba{\l}aban-Jaffe averaging kernel for the gauge field
\cite{balimja}.

The block Landau gauge is $\alpha =0$ . There exists
a formula which expresses
the fluctuation field propagator $\Gamma $ in the block Landau gauge in terms
of the propagator $\Gamma _{\kappa ,\alpha }$ for finite $\kappa $ (e.g.
$\kappa =1$ ) and arbitrary $\alpha $ \cite{balaban}.

\begin{eqnarray}
\Gamma _0=G-\frac 1\alpha G\,\partial \,{\cal R} \,\partial
^{\dagger}\,G-G\,C^{\dagger}\left( C\,G\,C^{\dagger}\right) ^{-1}\,C\,G
\end{eqnarray}
where $G=\Gamma _{1,\alpha }$.

The propagators $G$ and $\Gamma _0$ decay exponentially in coordinate
space with decay length one block lattice spacing. When the original
theory lives on a lattice, the proofs of reference \cite{balaban}
apply.
One can also
convince oneself of the fact by examining the behaviour in momentum space
near zero momentum. The explicit formulae given below can be used for
that.

The interpolation kernel ${\cal A}$ (which is independent of $\alpha $
as we know) can be expressed in terms of $G=\Gamma _{1,\alpha }$ as
well

\begin{eqnarray}
{\cal A\ }=G\,\,C\,^{\dagger}\left( CGC^{\dagger}\right) ^{-1}
\end{eqnarray}

Therefore, the block Landau gauge fluctuation field propagator $\Gamma
_0$ can also be written as

\begin{eqnarray}
\Gamma _0=G-\frac 1\alpha G\,\partial \, {\cal R}\,\partial ^{\dagger}
\,G-{\cal A}\,C\,G\,C^{\dagger}{\cal A}^{\dagger}
\end{eqnarray}
This resembles the formula for the scalar case. In place of the
propagator for the full theory, which does not exist because we have not
fixed the gauge completely, there appears the auxiliary quantity
$G=\Gamma _{1,\alpha } $. The expression for $\Gamma _0$ is valid for
any choice of $\alpha $ in this auxiliary propagator.

\subsection{Perfect Lattice action for the Maxwell field at Temperature $T
 > 0$}
To go from temperature $T=0$ to finite $T$
  one must periodize the propagators and
interpolation kernels in time as in the scalar case.

\begin{eqnarray}
D=(-\Delta _1)^{-1}
\end{eqnarray}
with Fourier transform $\tilde D(p)$.
Then the free lattice photon propagator
in the 3-dimensional lattice theory which is obtained from the finite
temperature field theory is

\begin{eqnarray}
\widetilde{D}_T({\bf p})=\widetilde D({\bf p},0)
\end{eqnarray}
Similarly, the interpolation kernel

\begin{eqnarray}
\widetilde{{\cal A}}_T(l,{\bf p})=
\widetilde{{\cal A}}(l,{\bf p},0)
\end{eqnarray}
and the fluctuation field propagator
\begin{eqnarray}
\widetilde{\Gamma}_{0T}(l,{\bf p},l^{^{\prime}})
=\widetilde{\Gamma}_0(l,{\bf p},0,l^{^{\prime}}).
\end{eqnarray}

As in the scalar case , the discrete variables $l,l^{^{\prime }}$ remain
4-dimensional. Think of them as referring to the original 4-dimensional
theory.

\section{Scalar electrodynamics}
\label{secqed}

Now we are prepared to deal with scalar electrodynamics.
The definitions for the gauge fields can be taken literally
from the preceding section.
Since the Higgs field is not invariant under a gauge transformation we
cannot simply use the results of section 3.
So it is the Higgs field we are concerned with in this section.

\subsection{Block spin transformation for the Higgs field at zero
temperature}
\label{subsechiggs}

Given a Higgs field $\phi(z)$, we wish to define a block Higgs field
$\Phi(x)$ on the lattice $\Lambda $ in a gauge covariant way. The
blocking procedure for the $\varphi ^4$-theory cannot be used as it
stands because it is not gauge covariant. In order to maintain gauge
covariance, one must use averaging kernels depending on the gauge
field $a.$

We denote the averaging operator for the Higgs field by $C^H(a)$

\begin{eqnarray}
\Phi =\,C^H(a)\,\phi
\end{eqnarray}

\begin{eqnarray}
\mbox{i.e.}\,\,\,\,\,\,\,\,\,\,\,\,\,\,\,\,\,\,\,\,\,\,\,\,\,\,\,\,
\,\,\, \,\,\,\,\,\,\Phi (x)=\int_{z\in x}dz\,\,\,C^H(a|\,x,z)\,\,\phi
(z)
\end{eqnarray}
The adjoint kernel $C^{H\dagger}(a|z,x)=\overline{C}^H(a|x,z)$

In the $\varphi ^4$-theory we used an averaging kernel which was
constant on blocks and vanished outside. A constant is the lowest
eigenvector of the Laplacian with Neumann boundary conditions on block
boundaries.

The natural generalization to the gauge covariant situation is as
follows
\footnote{Here we deviate from the work of Ba{\l}aban, Jaffe and
  Imbrie \cite{balimja}}.
Let $-\Delta
_a=\left( \partial -a\right) ^{\dagger}\left( \partial -a\right) $
denote the
covariant Laplacian and $-\Delta _a^{N,x}$ the covariant Laplacian with
Neumann boundary conditions on the block boundary of $x$. We demand that

\begin{eqnarray}
-\Delta _a^{N,x}C^{H\dagger}(a|z,x)=
\varepsilon _0(a|x)\;C^{H\dagger}(a|z,x)
\label{eigen}
\end{eqnarray}
and

\begin{eqnarray}
C^{H\dagger}(a|z,x)=0\,\,\,\,\,\,\,\,\,\,\,\,\,\,\,\,\,\,\,\,\,\,\,\,\,\,\,
\,\, \mbox{\thinspace \thinspace for\thinspace \thinspace \thinspace }
\,\,\,\,\,\,\,\,\,\,\,\,\,z\notin x
\end{eqnarray}
where $\varepsilon _0(a|x)$ is the lowest eigenvalue of $-\Delta
_a^{N,x}.$ In addition we impose the normalization condition

\begin{eqnarray}
C^HC^{H\dagger}=1
\end{eqnarray}
This leaves the freedom of multiplying $C^H$ with an $a$-dependent phase factor

\begin{eqnarray}
C^{H\dagger}(a|z,x)\longrightarrow C^{H\dagger}(a|z,x)\,\,\,\eta \,(a|x)
\end{eqnarray}

It follows from the gauge covariance of the eigenvalue problem (\ref{eigen})
that under a gauge transformation

\begin{eqnarray}
C^H(a^\lambda )\,\phi ^\lambda =\left( C^H(a)\,\phi \right) ^\Lambda
\label{ginv}
\end{eqnarray}
with some gauge transformation $\Lambda $ on the lattice which depends
on $ \lambda $ and on the choice of conventions to fix $C^H$ uniquely.
The freedom in the choice of an
$a$-dependent phase factor may be exploited to demand that

\begin{eqnarray}
\Lambda (x)=C\,\,\lambda (x)
\end{eqnarray}
where $C$ is the scalar averaging kernel introduced in subsection
\ref{subsecblock}.
Eq. (\ref{ginv}) parallels the gauge covariance property of the
blocking procedure for the gauge field.

One can compute the averaging kernel $C^H(a)$ as a solution of the
eigenvalue equation (\ref{eigen}) by standard quantum mechanical
perturbation theory.
A prototype of such a computation is found in reference \cite{palma}.

\subsection{The perfect action of scalar electrodynamics: Definition}
\label{subsecdefin}

The perfect action is defined by
\begin{eqnarray}
\exp \left( -S_{eff}(\Phi , {\bf A}) \right) &=& \int Da \, \int D\phi
\, \delta _{Ax}(a)\,\delta \left( C\,a - {\bf A}\right) \,\,
\delta _\kappa \left( C^H(a)\phi -\Phi \right)
\nonumber \\
& & \exp \left( -S_M(a)-\frac 12\left( \phi ,-\Delta _a\phi
\right) -V\left( \phi \right) \right)
\end{eqnarray}
where

\begin{eqnarray}
V\left( \phi \right) =\int dz\,\left[ \frac 12m_0^2\,\,\phi \,\phi
^{*}+\frac \lambda {4!}\left( \phi \,\phi ^{*}\right) ^2\right]
\end{eqnarray}
To evaluate it perturbatively, we proceed similarly as in
$ \varphi ^4$-theory and pure abelian gauge theory.

We prefer to use a Gaussian $\delta _\kappa \left( C^H(a)\phi -\Phi
\right) $ in place of a $\delta $-function.

\subsection{Interpolation kernel for the Higgs field and Higgs
  fluctuation propagator}
\label{subsecinter}

The interpolation kernel ${\cal A}^H$ for the Higgs field will also
depend on the gauge field $a$ because the averaging kernel does. In
order to split the kinetic term for the Higgs field, we impose the usual
demand that $\Psi = $ ${\cal A}^H\,\,\Phi \,$ minimizes $\left(\Psi
,-\Delta _a\Psi \right) $ subject to the constraint that $C^H(a)\Psi
=\Phi .$ This happens if

\begin{eqnarray}
\Delta _a{\cal A}^H(a|z,x)=\int_{y\in \Lambda }C^{H\dagger}(a|\,\,z,y)\Delta
_{eff}(a|\,\,y,x)
\end{eqnarray}
for some choice of $\Delta _{eff}$, and if

\begin{eqnarray}
C^H(a)\,{\cal A}^H(a)=1
\end{eqnarray}
The last condition implies that

\begin{eqnarray}
(\Delta_{eff}(a))^{-1}=C^H(a)\,\,\Delta^{-1}_aC^{H \dagger}(a)
\end{eqnarray}
similarly as before (where we used the notation $u^{-1}$ in place of
$\Delta_{eff}$).

If we make a shift of the Higgs field

\begin{eqnarray}
\phi ={\cal A}^H\Phi +\zeta
\end{eqnarray}
The fluctuation field propagator for the Higgs field is
\begin{eqnarray}
\Gamma ^H(z,w)=(-\Delta _a+ \kappa \,\,C^{H\dagger}C^H)^{-1}(z,w)
\end{eqnarray}
In the limit $\kappa \rightarrow \infty $ it becomes
\begin{eqnarray}
\Gamma ^H=v^H-{\cal A}^HC^Hv^HC^{H\dagger}{\cal A}^{H\dagger}
\end{eqnarray}
where $v^H=-\Delta _a^{-1}$ is the full gauge covariant free massless
propagator for the Higgs field.

The interpolation operator and the gauge
covariant fluctuation field propagator can also be computed by standard
quantum mechanical perturbation theory.

\subsection{The perfect action of scalar electrodynamics: Representation
as a functional integral in block Landau gauge}
\label{subseclandau}

We use again the Faddeev-Popov trick to convert from block axial gauge
to block Landau or $\alpha $-gauge. This is done in exactly the same way
as for the pure abelian gauge field theory.

The essential point which makes this possible is that both $\delta
$-functions

\begin{eqnarray}
\delta \left( C\,a- {\bf A}\right) \delta_{\kappa} (C^H(a)\phi -\Phi )
\end{eqnarray}
are invariant under gauge transformations $a\rightarrow a^\lambda
,\,\,\,\,\,\phi \rightarrow \phi ^\lambda $ which obey the constraint $
C\lambda =0.$ Proceeding as before, we obtain

\begin{eqnarray}
\exp \left( -S_{eff}(\Phi , {\bf A})\right) &=& \int Da \, \int D\phi
\, \delta \left( C\,a- {\bf A}\right)
\,\delta_{\kappa} (C^H(a)\phi -\Phi )
\nonumber \\
& & \exp \left( -S_M(a)-S_{gf}(a)-\frac 12\left( \phi ,-\Delta
_a\phi \right) -V(\phi )\right)
\end{eqnarray}
where

\begin{eqnarray}
S_{gf}(a)=\frac 1{2\alpha} \left( \partial^{\dagger}
\,a,\,{\cal R}\,\partial^{\dagger} \,\,a\right)
\end{eqnarray}
with the same ${\cal R}$ as for the pure abelian gauge field.

Now we shift the fields as

\begin{eqnarray}
\phi = {\cal A}^H \Phi + \zeta
\nonumber \\
a={\cal A} {\bf A}+a^{\prime}
\end{eqnarray}
The expression for the perfect action becomes

\begin{eqnarray}
\lefteqn{
\exp \left( - S_{eff}(\Phi, {\bf A}) \right) =
\int Da^{\prime} \, \int D\zeta \, \delta (C a^{\prime})
\delta_\kappa (C^H(a) \zeta)}
\nonumber \\
& &\hspace{3.0cm} \exp \bigg [ - S_{M,eff}({\bf A})
- S_{gf}({\cal A} {\bf A})
- S_M(a^{\prime}) - S_{gf}(a^{\prime})
\nonumber \\
& &\hspace{2.0cm}- \frac 12 \left( \Phi, - \Delta_{eff}(a) \,\, \Phi \right)
- \frac 12 \left( \zeta, - \Delta_a \, \zeta \right)
-V \left( {\cal A}^H(a) \Phi +\zeta \right) \bigg ]
\end{eqnarray}
where $a={\cal A} {\bf A}+a^{^{\prime }}.$

We separate the terms of zeroth order in $a^{^{\prime }}$ and $\zeta .$

\begin{eqnarray}
\Delta _{eff}(a)=\Delta _{eff}({\cal A} {\bf A})+E\left(
a^{^{\prime}},
 {\bf A}\right)
\equiv \widetilde{\Delta }_{eff}( {\bf A})+E(a^{^{\prime }}, {\bf A})
\end{eqnarray}

\begin{eqnarray}
V({\cal A\,}^H(a)\Phi +\zeta )=U_{cl}\left(  {\bf A},\Phi \right)
+W( {\bf A},a^{^{\prime }},\Phi ,\zeta )
\end{eqnarray}
where

\begin{eqnarray}
U_{cl}( {\bf A},\Phi )=V({\cal A}^H({\cal A} {\bf A})\Phi )
\end{eqnarray}
The perfect action becomes

\begin{eqnarray}
S_{eff}(\Phi , {\bf A})=S_{M,eff}( {\bf A}) + S_{gf}({\cal A} {\bf A})
+U_{cl}( {\bf A},\Phi )
-\frac 12\left( \Phi ,-\widetilde{\Delta }_{eff}( {\bf A})\Phi
\right)
\nonumber \\
+\widetilde{S}_{eff}(\Phi, {\bf A})
\end{eqnarray}
with

\begin{eqnarray}
\exp \left( -\widetilde{S}_{eff}(\Phi , {\bf A})\right) &=&
\int Da^{\prime}
D\zeta \,\delta (Ca^{\prime }) \delta_\kappa \left( C^H({\cal A}
{\bf A}+a^{\prime})\zeta \right)
\nonumber \\
& & \exp \bigg\{ - S_M(a^{\prime}) -S_{gf}(a^{\prime})
-\frac 12 \left( \zeta ,-\Delta_a \zeta \right)
\nonumber \\
& & \quad \quad -\frac12 \left( \Phi ,E(a^{\prime }, {\bf A})\Phi \right)
-W( {\bf A},a^{\prime},\Phi ,\zeta ) \bigg \}
\end{eqnarray}

To zeroth order in the fluctuation field propagators $\widetilde{S}
_{eff}(\Phi , {\bf A})=0$ after subtracting a constant.

\subsection{Scalar electrodynamics at finite temperature}
\label{subsecscqedt}

We wish to adopt the considerations of the preceding section to finite
temperature $T$.

We introduce a block lattice $\Lambda $ as for the $\phi ^4$-theory and
the Maxwell theory. The side length of the blocks in time direction is
$\beta $. As a result, the block lattice is 3-dimensional.

The blocking of the gauge field at finite temperature was discussed
before.

For the Higgs field, some new features appear, because the averaging and
interpolation kernels are gauge field dependent. For this reason, a
covariant momentum space description does not exist, and we cannot apply
exactly the same periodization procedure as for a purely scalar theory.
Nevertheless, the transition to finite temperature by periodization in
time is straightforward.
The formulae of the preceding section remain valid as they stand
when they are properly interpreted.

First of all, we must regard the blocks as coming equipped with periodic
boundary conditions in time direction. The Neumann boundary conditions
apply only to the boundaries of the block which remain, after periodic
boundary conditions in time direction are imposed.

Similarly, the defining equation for the interpolation kernel ${\cal
A}_T^H$ remains the same as for $T=0$ , except that periodic boundary
conditions in time direction are imposed, and the previously defined
averaging kernel $ C_T^H$ is to be used. This equation together with
the normalization condition $C_T^H{\cal A}_T^H=1$ also defines the
propagator
\begin{eqnarray}
u_{\mu \nu }(x,y)=-\Delta _{eff,\mu \nu }^{-1}(x,y)
\end{eqnarray}
for the blocked Higgs field. The fluctuation propagator $\Gamma _T^H$ is
obtained from these quantities as before and will also be temperature
dependent. All these quantities depend on the gauge field. This implies
a hidden $T$-dependence also for the block Higgs field, because the
boundary conditions for the gauge field are $T$-dependent.

The effective action is defined as before, with $C_T^H$ substituted for
$C^H$ , etc.. It lives on a 3-dimensional lattice $\Lambda $ . Because
of the anisotropy, the dependence of the resulting action on $ {\bf A}_0$ and
on $ {\bf A}_i$,$ \,\,\,\,\,\,i=1,2,3$ is different. The $ {\bf A}_0$
field behaves like an extra scalar field in the effective
3-dimensional theory.

\section*{Acknowledgement}
G.P. was partially supported by Fondecyt
$\#$ 1930067 and Dicyt $\#$ 049331PA.
U.K. wants to thank the DFG for financial support and USACH for hospitality.

\newpage

\begin{appendix}

\section{Partial integration of nonlocal terms in the effective
action to obtain a sum of local and irrelevant terms}
\label{local}

Consider a term which is quadratic in the field such as
\begin{eqnarray}
I_2(\Phi) = \int \int_{x_1,x_2} \rho_2 (x_2 - x_1) \Phi(x_2) \Phi(x_1)
\end{eqnarray}
We are interested in situations  where $\rho_2(x)$ falls off
exponentially with decay length one lattice spacing.
In this case the sum can be rewritten in the form
\begin{eqnarray}
I_2(\Phi) = \mu^2 \int_{x_1} \Phi(x_1)^2
+ z_{\mu \nu} \int_{x_1} \nabla_\mu \Phi(x_1) \nabla_\nu \Phi(x_1)
+ \mbox{irrelevant term}
\label{integral}
\end{eqnarray}
where the irrelevant term is of the form
\begin{eqnarray}
\gamma_{\mu \nu \rho \sigma} \int \int_{x_1,x_2}
\nabla_\mu \nabla_\nu \Phi(x_1)
\rho_2^{\prime} (x_2 - x_1) \nabla_\rho \nabla_\sigma \Phi(x_2)
\end{eqnarray}
$\rho_2^{\prime}$ also decays exponentially with distance $x_2-x_1$.
The coefficients are
\begin{eqnarray}
\mu^2 &=& \int_x \rho_2(x)  \\
z_{\mu \nu} &=& - \frac12 \int_x x_\mu x_\nu \rho_2(x)
\end{eqnarray}
Because of the exponential falloff of $\rho_2$, its Fourier transform
$\tilde \rho_2(p)$ is holomorphic in a strip.
Because of the presence of a lattice, it is a periodic and even
function of $p$.
Therefore
\begin{eqnarray}
\tilde \rho_2(p)= \mu^2 + z_{\mu \nu} \sin p_\mu \sin p_\nu
+ \gamma_{\mu \nu \rho \sigma}
\sin p_\mu \sin p_\nu \sin p_\rho \sin p_\sigma \,\, \tilde \rho^\prime_2(p)
\end{eqnarray}
where $\tilde \rho^\prime_2(p)$ is also holomorphic, periodic and even.
\begin{eqnarray}
\mu^2 &=& \tilde \rho_2(0)  \\
z_{\mu \nu} &=& \frac12 {\frac \partial {\partial p_\mu}}
{\frac \partial {\partial p_\nu}}
\tilde \rho_2(p)|_{p=0}
\end{eqnarray}
Inserting back one gets eq.(\ref{integral}).

\section{Dimensional reduction to a 3-dimensional theory in continuous
  space}
\label{dimred}

The method outlined in this paper can also be used to compute the
dimensionally reduced 3-dimensional theory in continuous space by
perturbation theory.
Its action is local to zeroth order, but will then start to develop
nonlocal terms of the form
\begin{eqnarray}
   \int \rho({\bf x_1},\ldots,{\bf x_n})
   \Phi({\bf x_1}) \cdots \Phi({\bf x_n})
\end{eqnarray}
where $\rho$ decays exponentially with the length of the shortest tree
on vertices ${\bf x_1},\ldots,{\bf x_n}$ with decay length $\frac
\beta {2\pi}$.

We outline briefly how this is seen.
The method had been used before in the work of M.\ Nie{\ss}en \cite{niessen}.
He used it to make d-dimensional quantum statistical systems more
palatable to a computer by discretizing time in the d+1-dimensional
functional integral representation of the system.

In our case, the 3-dimensional fields are time averages of the
4-dimensional ones.

\begin{eqnarray}
& &  \Phi({\bf r}) = \beta^{-1} \int_0^\beta dt \varphi({\bf r},t)
= \int dt^\prime d{\bf r}^\prime C({\bf r},{\bf r}^\prime,t^\prime)
\varphi({\bf r}^\prime,t^\prime)
\nonumber \\
& &C({\bf r},{\bf r}^\prime,t^\prime)= \beta^{-1} \delta({\bf r-r^\prime})
\end{eqnarray}
The block propagator $u({\bf r-r^\prime})$ is therefore given by

\begin{eqnarray}
u({\bf r}) =
\beta^{-1} u_{FT}({\bf r}) = \beta^{-1} \int_0^\beta dt \, v_T({\bf r},t) =
\beta^{-1} \int_{-\infty}^{\infty} dt \, v_0({\bf r},t)
\label{uft}
\end{eqnarray}
it depends on $\beta$ only through the overall factor (note that we did
not rescale the fields yet).

The interpolation operator is defined as usual, ${\cal A}= v_T
C^\dagger u^{-1}$.
Because $C$ is time independent, ${\cal A}$ comes out trivial.

\begin{eqnarray}
{\cal A}({\bf r^\prime},t^\prime,{\bf r}) = \beta^{-1}\delta({\bf r-r^\prime})
\end{eqnarray}
As a result the fluctuation propagator can also be simplified.
It is translation invariant both in ${\bf r}$ and $t$, and

\begin{eqnarray}
\Gamma_T({\bf r},t) = v_T({\bf r},t) -
\beta^{-1} \int_0^\beta dt^\prime v_T({\bf r},t^\prime)
\end{eqnarray}
The second term can be reexpressed using eq.(\ref{uft}).
It is seen that the Fouriertransform of $\Gamma_T$ is the same as of
$v_T$, except
for the absence of the zero Matsubara frequency mode.
Thus

\begin{eqnarray}
& & \Gamma_T({\bf r},t) =
\beta^{-1} \sum_{0\neq n\in Z} \int \frac {d^3{\bf p}} {(2\pi)^3}
e^{-i{\bf px} -i\omega_n t} (\omega_n^2 + {\bf p}^2 + m^2)^{-1}
\nonumber \\
& & \omega_n= 2\pi n \beta^{-1}
\end{eqnarray}
The decay in ${\bf r}$ can be read off the singularities of the
integrand on the imaginary $p$-axis ($p=|{\bf p}|$).
The closest singularity is at $p={\scriptstyle{+\atop -}} i\,2\pi \beta^{-1}$.
Therefore there is exponential decay with decay length $\frac \beta
{2\pi}$.

$\Gamma_T$ appears as free propagator in the perturbative expansion of
the 3-dimensional action.
Its decay properties will govern the nonlocalities of the resulting
action to all orders of perturbation theory.
This is familiar from the work of Gawedzki and Kupiainen.

\end{appendix}

\end{document}